\documentstyle[osa,manuscript]{revtex}

\begin{document}
\title{Superconducting Gap vs. Wave Vector: Evidence for Hot Regions 
on the Fermi Surface}
\author{R. Gatt, S. Christensen, B. Frazer, Y. Hirai, T. Schmauder, R. J. Kelley, M. Onellion}
\address{Physics Department, University of Wisconsin-Madison, 1150 University 
Avenue, Madison, WI 53706}
\author{I. Vobornik, L. Perfetti, G. Margaritondo}
\address{Institut de Physique Appliqu\'{e}e, \'{E}cole Polytechnique 
F\'{e}d\'{e}rale 
Lausanne, CH-1015 Lausanne, Switzerland}
\author{A. Morawski, T. Lada, A. Paszewin}
\address{High Pressure Research Center Unipress, Polish Academy of 
Sciences, Warsaw, Poland}
\author{C. Kendziora}
\address{Naval Research Laboratory, Washington, D.C. 20011, USA}
\maketitle
\begin{abstract}                
 We have used angular resolved photoemission to measure the angular
  dependence  of the superconducting gap in highly overdoped Bi2212
   (Tc=65K). While the node at 45 degrees is conserved, we find 
   substantial deviation from a first order d-wave dependence away 
   from the node. The pairing susceptibility is peaked at special 
   regions on the Fermi surface. Comparing these results with a detailed 
   mapping of the Fermi surface we performed, we could measure the 
   extension and location of these hot regions. We find the hot regions 
   to be evenly spread about the nominal locations of hot spots. 
   The decrease of the gap amplitude away from these hot regions 
   follows very closely theoretical calculations within the spin 
   fluctuation approach. These results strongly suggest that the 
   pairing susceptibility is peaked at $Q=(\pi,\pi)$.
\end{abstract}
\section{Text}               
The mechanism of high temperature superconductivity is 
one of the most important current problems
 of condensed matter physics, and a detailed description 
 of the interaction which leads to pairing is still lacking.
 In common with most metals, the cuprate normal state exhibits 
 a Fermi surface\cite{phys report}.  
 Phase sensitive experiments,\cite{Wollman,Tsuei} indicate
  that the superconducting gap 
 (the order parameter of this second order phase transition) exhibits
  a change of sign around the Fermi surface. 
 The most popular such unconventional order parameter (symmetry different 
 from the crystal lattice) is a "simple d-wave," 
 in which the gap varies with angle around the Fermi surface as 
$\Delta(\Phi)=\Delta(0)\cos(2\Phi)$.
  Here $\Phi=45\deg$ corresponds to the $(\pi,\pi)$ direction in 
  reciprocal space. 
  Such an angular dependence is  observed on optimally doped 
  samples\cite{Shen,Ding}. 
  Here we report on $\Delta(\Phi)$ measurements on highly overdoped 
  samples.  
  Our main result is that the gap versus angle data deviate from a 
  simple d-wave order parameter. 
   Combined with a detailed mapping of the Fermi surface, these results
    indicate that particular
    regions ("hot regions") of the Fermi surface show very high pairing
     susceptibility. 
    This in turn provides insight into the superconducting pairing mechanism. \\
A general discussion on the electronic structure of the cuprates was given 
by Shen and Schrieffer\cite{S&S}. They have stressed the difference in the
 line shape and doping dependence
 of the spectral function along the $\Gamma - M$ direction (parallel to the 
 Cu-O bond) vs. along the $\Gamma - Y$ direction
  (parallel to the Cu-Cu direction). They suggested that such behavior could
   arise from electronic scattering 
  peaked at $Q=(\pi,\pi)$. They further suggested this scattering mechanism to 
  evolve into pairing susceptibility peaked 
  at $Q=(\pi,\pi)$ at low temperatures. In this work, we provide evidence for
   such a behavior. \\
  A feature observed in the underdoped cuprates is the pseudo-gap\cite{7Ding,8Shen}
   which manifests 
  itself as a reduction of the spectral intensity at the Fermi level and a shift
   of spectral weight to higher 
  frequencies. Its origin is believed to be related to incoherent pairing susceptibility
   existing above 
  $T_c$\cite{Ding-FS,Chub-Lead-edge}. The value of the pseudo-gap decreases 
  monotonically with temperature and continuously 
  transforms into the value of the superconducting gap. This makes
   it difficult to distinguish between the true 
  superconducting gap and the pseudo-gap\cite{MO}. 
  For high enough doping range there is no evidence of a 
  pseudo-gap forming\cite{Tallon} and the leading edge gap closes at 
  $T_c$.\cite{Loeser} 
  We therefore concentrate in our studies on the 
  highly overdoped cuprates. Another advantage in studying the highly overdoped 
  range is the general trend toward Fermi 
  liquid behavior with increased doping\cite{Misra}. This trend yields much 
  sharper spectral features compared with the
   underdoped region. As a result we were able to reduce our error bars and 
   detect fine details on which we report here. \\
   
  In a recent work we reported the existence of a node 
  in the gap amplitude in highly overdoped Bi2212 samples,
   consistent with a d-wave order parameter\cite{10Gatt-preprint}.
    We found that to obtain results reproducible 
   to gap sizes of ±1meV, extreme care had to be taken to avoid even
    very small amounts of twinning or inter-growth in our
    crystals. Further, due to the angle-resolved photoemission system
     we used, we found that by using large, 
    (typically $5\times 5\times 0.1 mm^3$ ) single crystals, we reduced the background
    in our spectra enough to distinguish between
     very small and zero values of the gap. We attribute the earlier reports 
     by some of us\cite{Kelley Science 96} that the gap in the 
     $(\pi,\pi)$
      direction is non-zero as due primarily to small amounts of material 
      imperfections. The high quality of the crystals 
      was needed to assure accurate alignment and to observe fine details
       such as superlattice bands. This last observation
       was important to assure measurements in the Y quadrant of the Brillouin
        zone, thus avoiding side effects rising from
        the Bi-O superlattice modulation\cite{Norman SL}. 

We used several experimental methods to characterize our samples, 
including polarized light microscopy, 
x-ray Laue diffraction, x-ray $\theta -2\theta$ diffraction and 
scanning electron microscopy with Kikuchi pattern analysis. 
We extracted and characterized samples from batches grown by self flux 
methods\cite{Han}, using our previous experience 
in crystal growth and characterization\cite {Gatt physica C}. To obtain 
overdoped (high carrier concentration) samples, 
we used two methods.
In one, we annealed the crystals in an atmosphere of 20\% oxygen and 80\% argon
 at a total pressure of 10 kbar. In the other,
 we sealed the samples in liquid oxygen and annealed them at 400C. Our data
  were obtained on the largest of these
  single crystals, up to $5\times5\times0.1 mm^3$ in size. The value of $T_c$ varied 
  between 60-65K. 
The angle resolved measurements were performed using a Scienta hemispherical
 energy analyzer
 with a mean radius of 300 mm. The measured full-width half maximum energy 
 resolution was 13-15 meV. 
 The light source was a helium discharge lamp providing photons of 21.22 eV
  photon energy.  
 The base pressure of the angle-resolved photoemission chamber was below 
 $2\times10^{-10}$ Torr and the sample
  was transferred from a load lock chamber, cooled to 11K, and a fresh surface
   exposed by cleaving the sample in situ. 
   The low temperature assured us that the sample did not lose oxygen. 
    The analyzer has a finite angular resolution of 
    $4.5\deg$ , in terms of the angle $(\phi)$ with respect to the Y-M direction 
   in reciprocal space.  \\
   Figure 1 is a comparison between the leading edge of the spectral function
    and the Fermi edge obtained from 
   a freshly evaporated Ag film. In figure 1a, the two edges are coinciding, 
   yielding a node in the superconducting 
   gap within our sensitivity limit. Fig. 1b presents a gap of 1meV measured
    as the distance between the leading edge 
   of the spectral function and the Fermi edge at the point of half maximum 
   intensity. The figure demonstrates that the 
   uncertainty in estimating the gap size is no more than $\pm 1$ meV. 
   Fig. 1c presents the gap at the $(0,\pi)$ point and Fig. 1d at 
   $\phi=15 deg.$ on the Fermi surface. \\
   
   Fig. 2 presents the  data collected on a total of 4 such crystals. The angular
    dependence of the superconducting gap
    with respect to the $(\pi,\pi)-(0,\pi)$ direction of the Brillouin zone is
     shown. The solid diamonds are measurements 
    on the Fermi surface. The open diamonds are measurements away from the Fermi
     surface. The behavior close to the 
    node follows very closely $ \cos(2\phi)$ dependence, compatible with a simple,
     first order d-wave behavior, represented 
    here by the dotted black line. There are important deviations from this 
    dependence away from the node. It is seen 
    clearly from the figure that the superconducting gap increases sharply in 
    amplitude at special regions on the Fermi 
    surface. Around 30 degrees, the gap amplitude increases sharply. This 
    increase saturates around 22 degrees.  \\
    
    A major ingredient of the spin fluctuation 
    approach\cite{Moriya,MP} to the problem of cuprate
     superconductivity is the existence of an interaction susceptibility
      peaked at a wave vector $Q=(\pi,\pi)$. 
     This leads to special points on the Fermi surface (so called hot spots) 
     which are connected by that 
     displacement in k-space. To explore the possible connection between the
      angular dependence of the 
     superconducting gap and the detailed shape of the Fermi surface, we 
     present in Fig. 3 a detailed mapping 
     of the Fermi surface for this high doping range. The mapping was done 
     mainly at the Synchrotron Radiation 
     Center in Wisconsin, and fine details were compared and completed in
      Lausanne. At Wisconsin we have used a 
     VSW hemispherical electron energy analyzer with a total resolution of
      35 meV. Recently we have repeated these
      measurements using a Scienta 200 energy analyzer on the PGM beamline 
      at Wisconsin. We have measured cuts along
       high symmetry directions and more than 20 cuts along other directions
       , most of them in parallel to the 
       $\Gamma-M$ direction. For obtaining the Fermi surface crossing points we have
        used the criteria developed by 
       Ding et. al\cite{Ding-FS}. An example for such cuts is presented 
       in the inset of figure 3.  \\
       
       Fig. 4 combines the results presented in
        figures 2 and 3. The nominal location of 
        hot spots is given
        as the intersection of the dashed square
         with $Q=(\pi,\pi)$ as its side and the Fermi surface.
          As can be 
        seen from Fig. 4 there are exactly eight such points on 
        the Fermi surface (filled circles). These so called hot spots 
        are the basis of numerous theoretical works within the
         spin fluctuation 
        approach\cite{hot ref} and there 
        were several attempts to observe them 
        experimentally\cite{Aebi,Bianconi}. However It was shown 
        that the features observed experimentally could 
        be naturally described as a structural effect, a result of 
        the superlattice modulation in 
        Bi2212\cite{Ding SL,Mesot hot}. The observation presented 
        in Fig. 2 is a direct measurement of 
        the true superconducting gap ( and not of a pseudogap). 
        It is seen clearly from the figure that the pairing 
        susceptibility is peaked at isolated regions on the Fermi surface. 
        By performing measurements in 
        the Y quadrant of the Brillouin zone, we distinguish between the
         main band crossing and the side bands 
        which are the result of the superlattice modulations. Our measurement 
        yields a value of $K_{co} = (0.95 \pi/a, 0.12\pi)$, 
        in polar units $(\rho,\phi)$ with respect to the Y point and the Y-M line, 
        as a cutoff vector for this region of high 
        pairing susceptibility on the Fermi surface. The construction of 
        fig. 4 yields $K_{hs} = (0.96 \pi/a, 0.06 \pi)$ for our 
        doping range. Notice that the measured $K_{hs}$ falls exactly in 
        the middle between $K_{co}$ and the zone boundary. We 
        emphasize that both $K_{hs}$ and $K_{co}$ are determined by the 
        data of Figs. 2 and 3, rather than a theoretical assumption.
         Recently, Abanov et al. developed a new procedure for calculating 
         $T_c$ in the scenario where the pairing
          susceptibility is highly peaked at a hot spot\cite{ACF}. The  solid  
          line in Fig. 2 is a one parameter 
          fit to their expression, describing the decrease in magnitude of the
           superconducting gap away from a hot spot.
           The fit yields a value of $0.9\pm 0.1 nm$ (about two unit cells) for the magnetic
           correlation length, which is as expected for this highly overdoped region of the phase 
           diagram.  \\

   In conclusion, we measured the angular dependence of the superconducting gap.
    We found substantial variation  
   from a first order d-wave behavior away from the node. We found that the pairing
    susceptibility is peaked at
    special regions on the Fermi surface. Using a detailed mapping of the Fermi 
    surface for this doping range, 
    we could measure the nominal hot spot location and the extension of our measured 
    hot regions. We have found 
    the hot regions to be evenly spread about the nominal locations of hot spots. 
    The decrease of the gap amplitude
     away from these hot spots follows very closely theoretical calculations within
      the spin fluctuation approach.
      From a fit to these calculations we have extracted the magnetic correlation 
      length and found it in excellent
      agreement with our highly overdoped range. \\
      
Note again, that the hot spots are not highly confined in k-space,
 but occupy considerable regions on the Fermi 
surface. This behavior can be described within the spin fluctuation approach 
as a result of the short magnetic 
correlation length in this doping range. Thus, as the magnetic correlation length
 decreases away from half filling, 
the pairing susceptibility acquires a finite broadness while still peaked 
at $Q=(\pi,\pi)$. Our results suggest that underdoped samples should 
have less broadening, with the hot regions becoming hot spots.  \\

Whatever is the pairing mechanism, our measurements show a high anisotropy in the
 angular dependence of the
 superconducting gap that is inconsistent with a simple d-wave picture (but still
  consistent with general 
 $B_{1g}$ symmetry). Our measurements show clearly a steep increase of the 
 superconducting gap at hot regions on the Fermi
  surface implying their overall importance in inducing superconductivity. 
  Both the sharp
   increase of the gap amplitude, and 
  the measured range on the Fermi surface for which this increase is observed, 
  suggest very strongly that the electronic 
  scattering that is the precursor for pairing in the high temperature 
  superconductors is peaked at $Q=(\pi,\pi)$.

ACKNOWLEDGEMENTS. We benefited from discussions with P. Chaudhari, S. Rast, 
A. Chubukov, M. Rzchowski, L. Forro, M. Grioni, A. Leggett, M. Weger, C.C. Tsuei, 
R. Joynt and I. Panas. Financial support was provided by the U.S. NSF and Fonds 
National Suisse. 

CORRESPONDENCE should be addressed to R.G.  (e-mail: Rafi@src.wisc.edu).

 \begin{figure}  
 \caption{The superconducting gap as extracted from the difference between
  the leading edge of the spectral function and the Fermi edge obtained from the 
  spectrum of a freshly evaporated Ag film. (a)  zero gap value, the two edges are 
  coinciding, (b) 1meV gap value, as extracted from the difference of the two spectra 
  at the point of half maximum intensity. The inset shows the extended energy scale in
   both figures. (c), (d) The shift of the leading edge at the 
   $(0,\pi)$ point and the
    point c on the Fermi surface as indicated by the inset in both figures.}
 \caption{The angular dependence of the superconducting gap in highly
  overdoped (Tc=65K) Bi2212. The dashed line is a fit to a first order d-wave gap 
  $(cos2\theta)$. The solid line marked ACF is a fit to Abanov et. al. expression for the
   decrease in the gap value away from a hot spot.}
  \caption{The Fermi surface of highly overdoped Bi2212. Open circles: 
  measurement points. Filled circles: main Fermi surface crossing points. Gray diamonds: 
  superlattice band crossing points. Open squares were obtained by symmetry operations. 
  The solid line is a smooth fit to the data, used in figure 4. Inset: Energy dispersive
   curves close to the main Fermi surface crossing points along directions a and b as
    indicated in the figure. Both directions are perpendicular to 
    $\Gamma-M$ direction.}
    \caption{The location of hot regions on the Fermi surface. These regions of
     high gap amplitude are extended from the zone boundary up to 
     $K_{co}$, the cutoff vector 
     for which the gap amplitude falls sharply to a first order d-wave 
     value. $K_{hs}$ are the
      locations of nominal hot spots, obtained from the intersection of the Fermi surface
       and the square with $Q=(\pi,\pi)$ as its side. Note that $K_{hs}$ falls exactly in the
        center of the hot region.}      
 \end{figure}
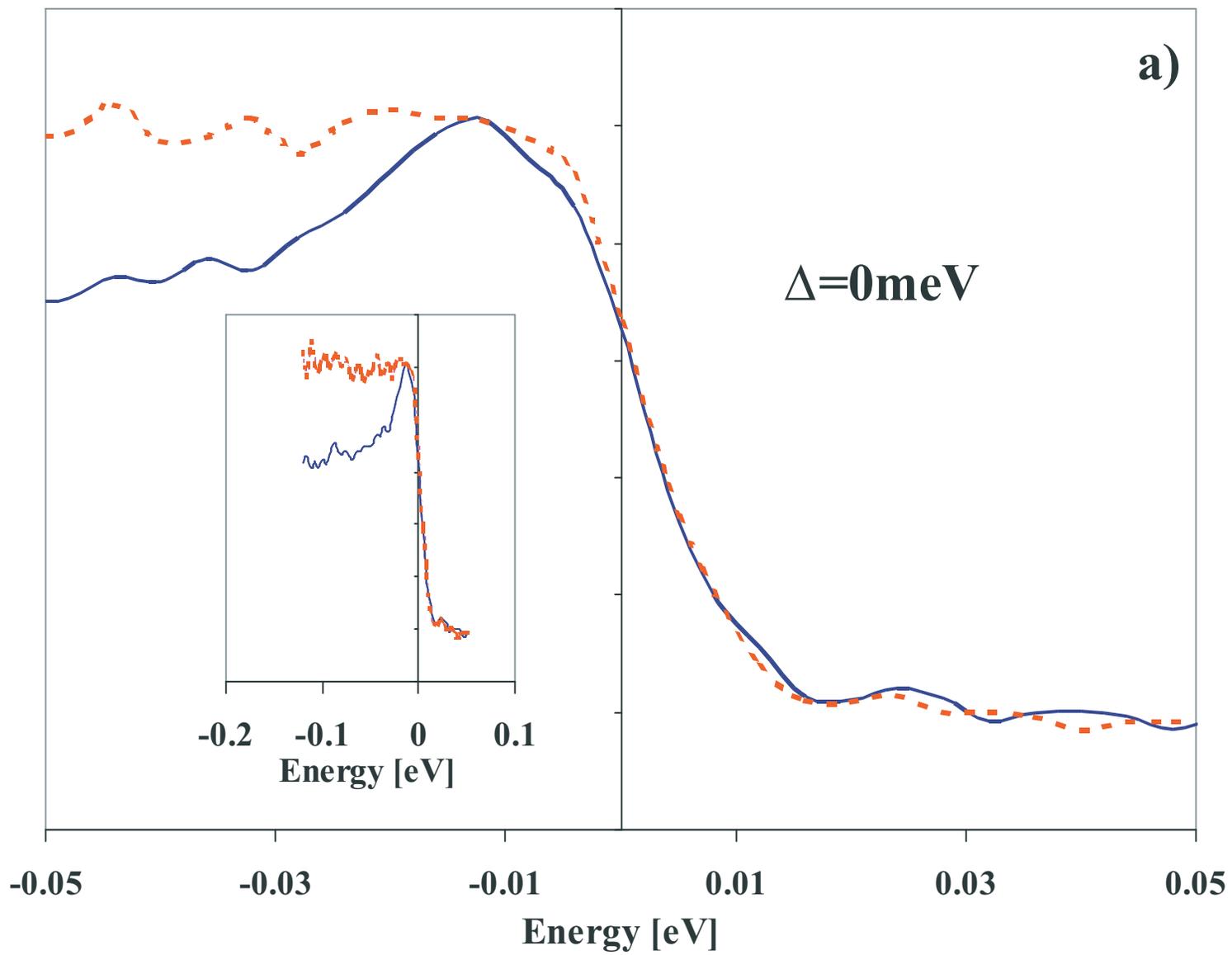
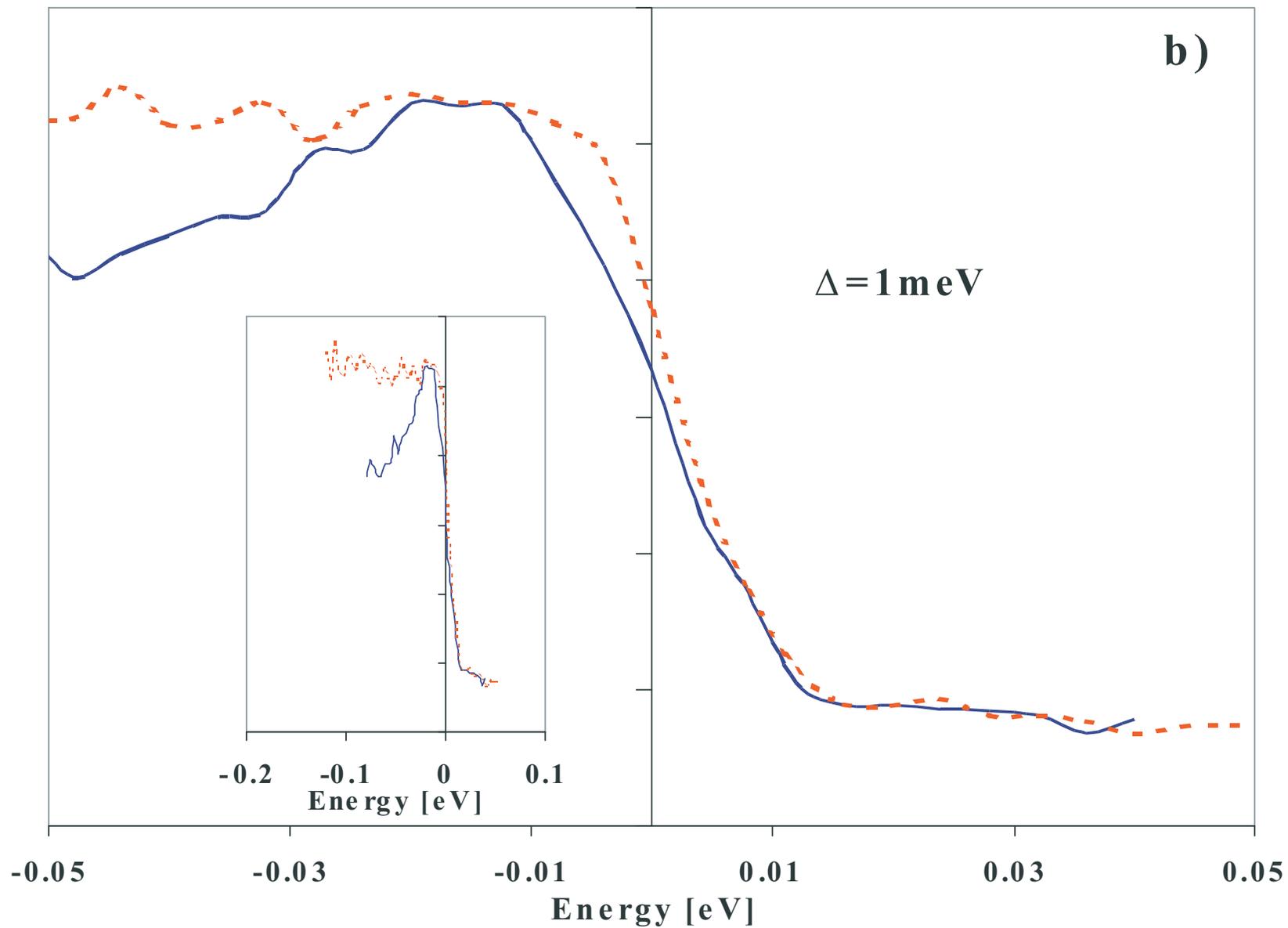
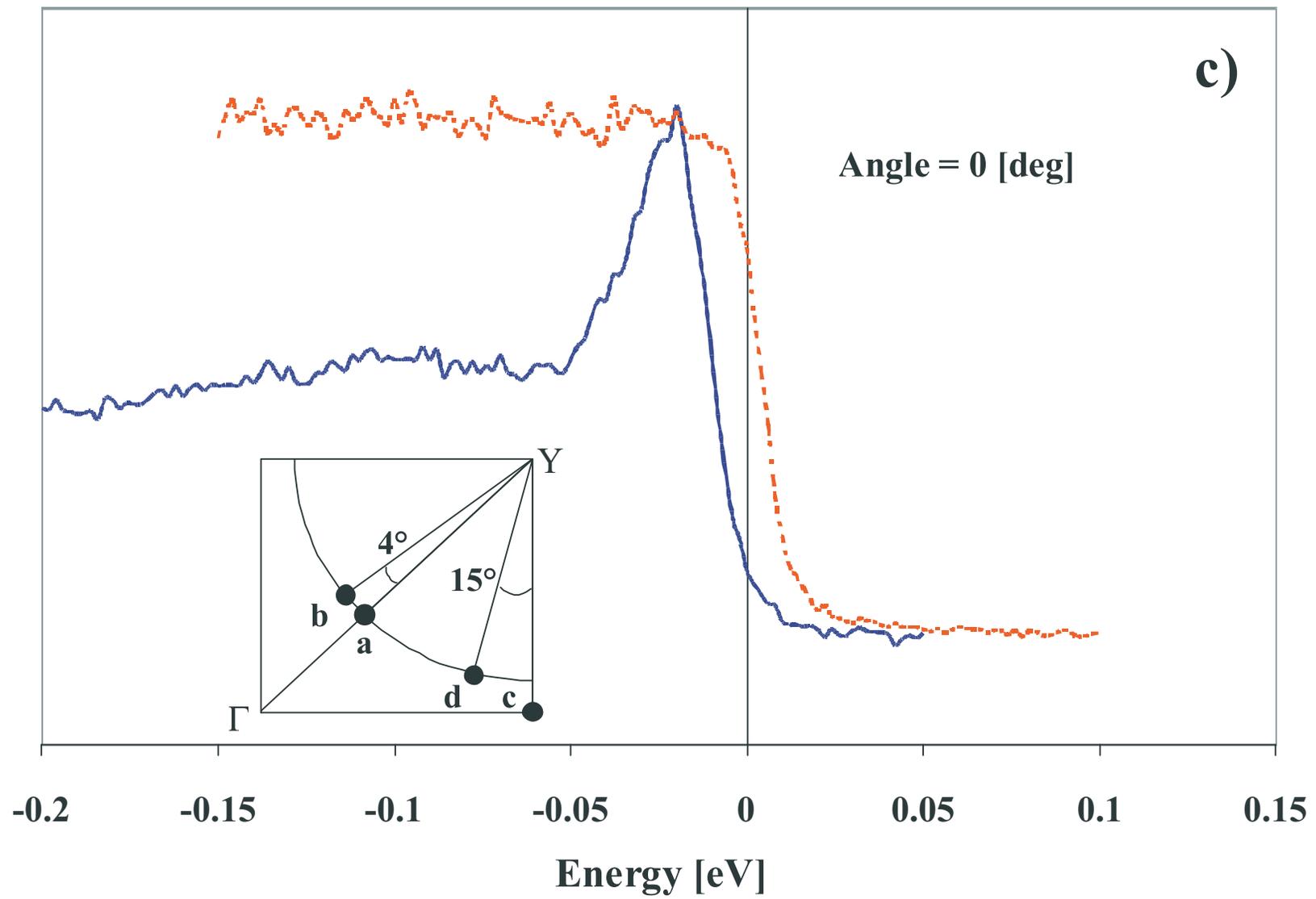
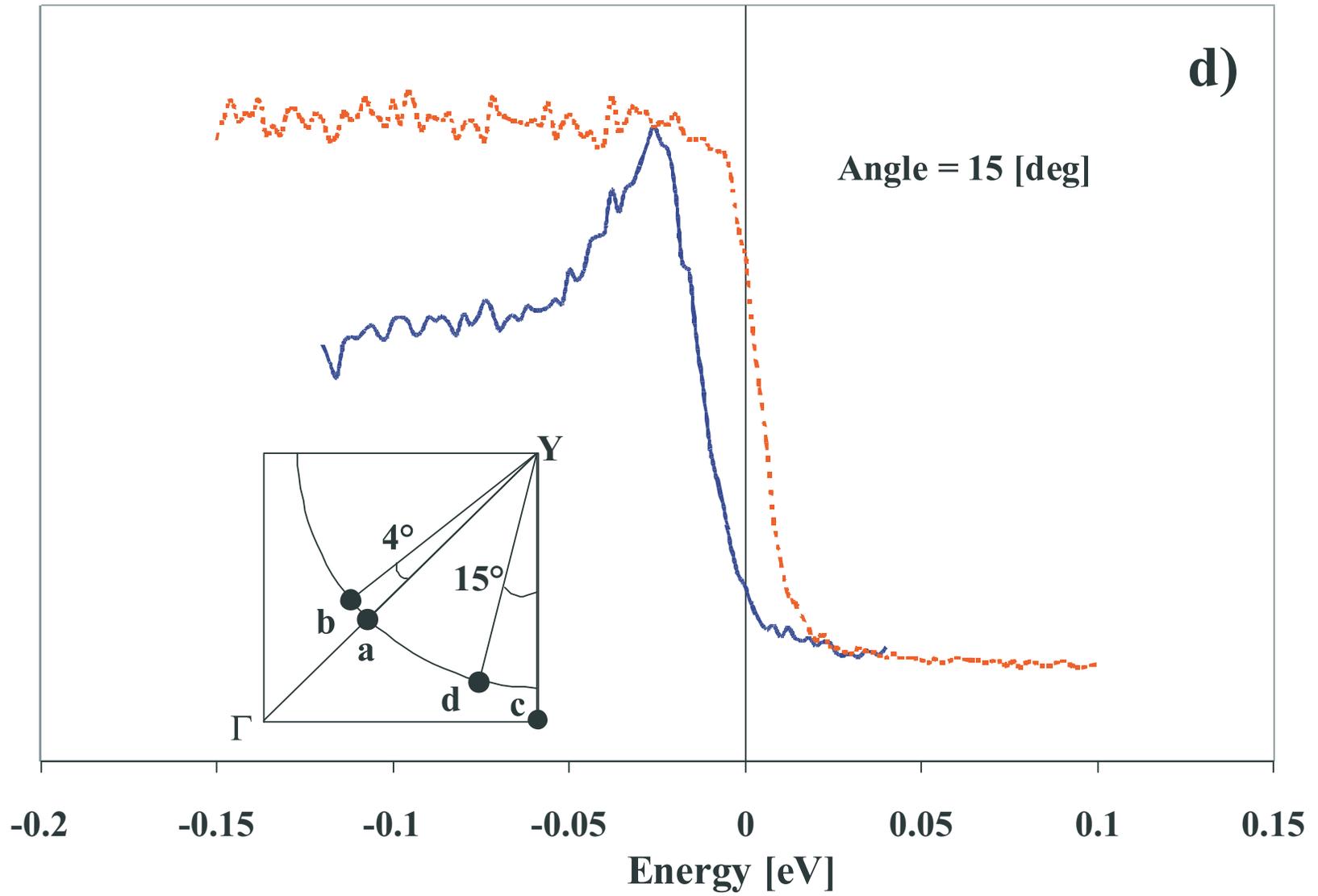
\end{document}